\documentclass[a4paper,11pt]{nveart}
\usepackage{graphics,pifont,amsmath,amssymb}

\begin{document}

\begin{frontmatter}
\title{In-event background and signal reconstruction for\\
       two-photon invariant-mass analyses}

\author{Nick van Eijndhoven \and Wouter Wetzels}
\address{Department of Subatomic Physics, Utrecht University/NIKHEF\\
         P.O. Box 80.000, NL-3508 TA Utrecht, The Netherlands}

\begin{abstract}
A method is presented for the reconstruction of both the background and signal in
invariant-mass analyses for two-photon decays.
The procedure does not make use of event mixing techniques and as such is based
exclusively on an event-by-event analysis.
Consequently, topological correlations of the event (e.g. jet structures) are
automatically taken into account.
By means of the decay process $\pi^{0} \rightarrow \gamma \gamma$ it will be demonstrated
how the procedure allows for determination of the $\pi^{0}$ yield from the observed
decay photons.
\end{abstract}

\end{frontmatter}

\section{Introduction}
In particle physics experiments some particles have such a short lifetime that direct detection
is impossible.
In these cases the decay products have to be used to reconstruct the parent-particle distributions.
For example, an invariant-mass analysis allows to determine the yield and also the transverse-momentum
distribution of the parent particles \cite{wa98pt}.

In invariant-mass analyses it is essential to separate the
invariant-mass peak from the combinatorial background. Using an event
mixing procedure an approximation for this combinatorial background can be
obtained. The invariant-mass peak is then reconstructed from the total invariant-mass
distribution by subtracting this background.
In order to construct the background by means of such an event mixing procedure, 
particles originating from different events have to be combined in pairs. 
Differences in the characteristics of the events used in the mixing procedure will cause the 
constructed background to deviate from the actual combinatorial background. 
The degree of deviation from the actual background depends on the position 
correlations of the decay particles.
For instance, in the determination of the transverse-momentum distribution of the parent particles, 
this background has to be constructed for various transverse-momentum values. 
The influence of jet structures might be restricted to a certain transverse-momentum range,
which implies that the correlations of the decay particles are related to the transverse momentum
of the parent particles.
In this way a systematical effect is introduced which affects (the slope of) the transverse-momentum
distribution.

From the above it is seen that it would be preferable to use a method that does not
require combination of particles from different events.
Such a method, which takes topological correlations of the particles within the event into account,
is presented here.
It is applied to the analysis of the decay process $\pi^{0} \rightarrow \gamma \gamma$
under circumstances matching those of Pb+Pb collisions at the CERN-SPS.
The datasets used for evaluation of our analysis method were obtained by means of 
a computer simulation that models the $\pi^{0}$ spectra as observed in heavy-ion
collisions at SPS energies \cite{prl85}.\\
A realistic energy resolution has been implemented to mimic
the photon detection in an actual calorimeter system like the one of 
the WA98 experiment \cite{prl85}. The details of this simulation are given 
in section \ref{sec-compsim}.

\section{The invariant mass distribution}
Consider a heavy-ion collision in which a certain amount of $\pi^{0}$s is
produced. More than 98 percent of these $\pi^{0}$s will decay into a pair
of photons.
Because of their short lifetime ($c\tau=25.1$~nm), the $\pi^{0}$s can be regarded to decay
immediately after their creation at the interaction point. Under realistic experimental 
conditions only a fraction of the decay photons will be detected. The energy and
position of a photon $i$ are recorded in the detection system and will be
denoted by $E_{i}$\ and $\vec{x}_{i}$, respectively. Consequently, the
four-momentum of each detected photon is determined.

The invariant mass $M_{inv}$ of a pair of photons with four-momenta $p_{1}$
and $p_{2}$ is computed in the following way~:

\begin{equation}
M_{inv}^{2}=(p_{1}+p_{2})^{2}=2(E_{1}E_{2}-\vec{p}_{1}\cdot\vec{p}_{2})
           =2E_{1}E_{2}(1-\cos(\vartheta_{12})) ~,
\end{equation}
where $\vartheta_{12}$ is the angle between the momenta of the two photons.

All possible combinations of two photons within an event will provide a distribution
consisting of the $\pi^{0}$ invariant-mass peak and a combinatorial background.
The latter being due to combinations of two photons which did not both originate
from the same $\pi^{0}$ decay.
The distribution obtained in this way will be called the total invariant-mass distribution
$\mathcal{T}$.\\
The combinations of energies and positions that constitute this $\mathcal{T}$ distribution are 
indicated below

\begin{eqnarray*}
&&E_{1}\vec{x}_{1} \otimes E_{2}\vec{x}_{2} \quad \text{(photon 1 combined with photon 2)},\\
&&E_{1}\vec{x}_{1} \otimes E_{3}\vec{x}_{3} ~, \\
&&\ldots \\
&&E_{N-1}\vec{x}_{N-1} \otimes E_{N}\vec{x}_{N} ~,
\end{eqnarray*}
in which a combination $E_{i}\vec{x}_{i} \otimes E_{j}\vec{x}_{j}$ yields an invariant mass
$M_{ij}=\sqrt{2E_{i}E_{j}(1-\cos(\vartheta_{ij}))}$.

\begin{figure}[htb]
\begin{center}
\resizebox{10cm}{!}{\includegraphics{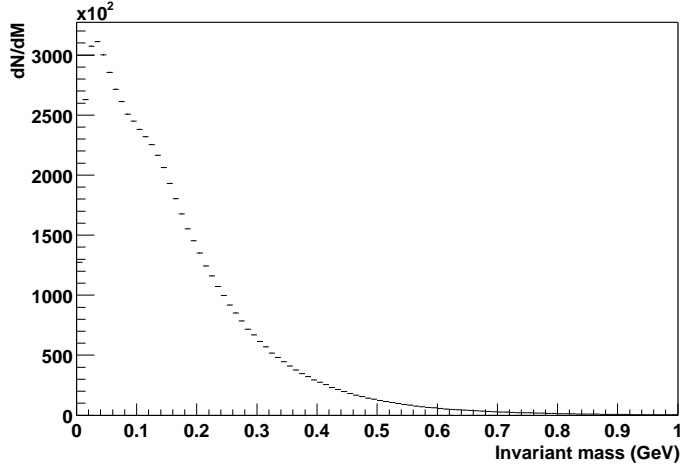}}
\end{center}
\caption{The total invariant-mass distribution $\mathcal{T}$ for the case of 50 $\pi^{0}$s/event.}
\label{fig:t50}
\end{figure}

An example of such a total invariant-mass distribution, resulting for events containing
50 $\pi^{0}$s each, is shown in fig.~\ref{fig:t50}.
Here the $\pi^{0}$ peak at about 135~MeV is seen positioned on top of a large combinatorial background.
Since the integral of the peak corresponds to the number of $\pi^{0}$s for which both decay photons were
detected, being the basic ingredient in the studies of $\pi^{0}$ spectra, it is essential to
separate the $\pi^{0}$ peak from the background.\\
This cannot be done with sufficient accuracy in a straightforward way, since many combinations
of photons that contribute to the background have an invariant mass close to the $\pi^{0}$ mass.
In addition, the width of the reconstructed $\pi^{0}$ peak is relatively large, due to the
limited energy and position resolution of the detection system, and the shape of the combinatorial
background is unknown.\\
If an accurate prediction of the (shape of the) background distribution could be obtained,
the $\pi^{0}$ peak could be retrieved from the total invariant-mass distribution $\mathcal{T}$
simply by subtracting this background.

\section{Background reconstruction via event mixing}
A distribution that accurately represents the combinatorial background can be obtained by
means of an event mixing method.
For each entry in the "event mixing" background distribution, two sets of photon momenta
(obtained from two different events) are used. One photon momentum is taken from the first of
these two sets and another is taken from the second. These two are then
combined to provide an entry in the invariant-mass distribution.\\
The combinations used in the case of mixing events $\rm{A}$ and $\rm{B}$, of which the numbers
of detected photons are $N_{\rm{A}}$ and $N_{\rm{B}}$ respectively, are given by~:

\begin{eqnarray*}
&&E_{1[\rm{A}]}\vec{x}_{1[\rm{A}]}\otimes E_{1[\rm{B}]}\vec{x}_{1[\rm{B}]} ,\\
&&E_{1[\rm{A}]}\vec{x}_{1[\rm{A}]}\otimes E_{2[\rm{B}]}\vec{x}_{2[\rm{B}]} ,\\
&&\ldots \\
&&E_{N_{\rm{A}}[\rm{A}]}\vec{x}_{N_{\rm{A}}[\rm{A}]}
  \otimes E_{N_{\rm{B}}[\rm{B}]}\vec{x}_{N_{\rm{B}}[\rm{B}]}.
\end{eqnarray*}

The invariant-mass distribution obtained in this way is expected to describe the actual
combinatorial-background distribution rather accurately.
No combinations of two photons originating from the same $\pi^{0}$ decay occur.
Therefore, the $\pi^{0}$ peak will be absent.\\
An imperfection of this method is the fact that the two events used for mixing will have different
position and energy distributions. In addition, for heavy-ion collisions also the centrality of the
two events is different.
The latter may be partly overcome by grouping the collisions into centrality classes and combining
only events that belong to the same class. Still, possible correlations in photon positions
within the original event will surely be affected when a second event is used in the
mixing procedure.\\
Since in general the $\pi^{0}$ peak is relatively small compared to the background, the error in
the derived number of parent particles due to these effects might become substantial, especially
in the case of constructing transverse-momentum spectra.

\section{Background and signal reconstruction via position swapping}
\label{sec-posswap}
Below we present an alternative method to reconstruct both the combinatorial
background and the invariant-mass peak signal of the parent particles.
The procedure is carried out within one and the same event and as such does not
require mixing of data from different events.\\
In order to perform this analysis a new distribution has to be constructed, to be
called the "position swapped" distribution $\mathcal{S}$.
For each entry in this distribution two photons are taken from a single event.
However, before computing the invariant mass of this combination, one of the two
photon positions is replaced by the position of a randomly chosen third
photon from the same event.
The combinations which constitute this $\mathcal{S}$ distribution are
given by $E_{i}\vec{x}_{i} \otimes E_{j}\vec{x}_{\alpha }$, where $i$ and $j$ indicate photon
$i$ and $j$ as before and $\alpha$ denotes the randomly chosen third photon ($\alpha \neq i \neq j$).

Unlike in the case of event mixing, only one event is used to compute each
entry. An important advantage is that when the photon positions are swapped
the global characteristics of the event are not modified, thus preserving possible 
topological correlations.\\
In the case of azimuthal symmetry and a significant energy--position correlation 
for the photons, it is recommended to divide the photon sample into categories 
characterised by the photon polar angle.
In this case the energy--position correlation can be maintained in the position-swapping process,
by only combining photons belonging to the same category.

To enhance the statistical significance, the number of entries can be increased by
allowing $\alpha $ to take multiple values.
Furthermore, it is recommended to replace both $\vec{x}_{i}$ and $\vec{x_{j}}$ subsequently
as described before to make sure that all of the photon positions occur with comparable statistical
weights. In order to extract the signal, the $\mathcal{S}$ distribution has to be normalised
properly, as outlined hereafter.

\begin{figure}[htb]
\begin{center}
\resizebox{10cm}{!}{\includegraphics{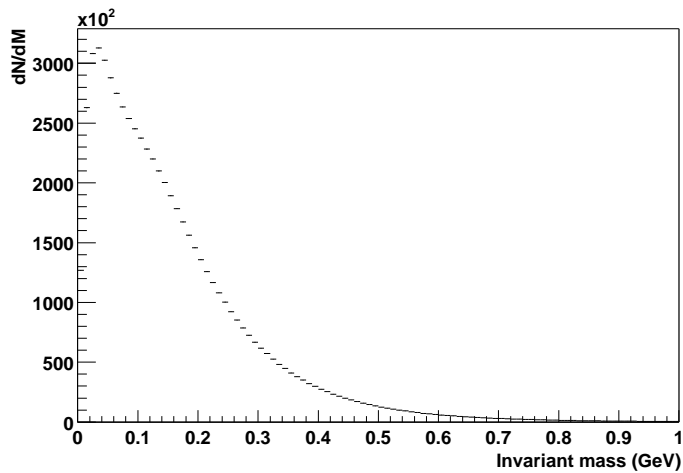}}
\end{center}
\caption{The background distribution $\mathcal{S}$ for the case of 50 $\pi^{0}$s/event.}
\label{fig:s50}
\end{figure}

An example of the resulting $\mathcal{S}$ distribution corresponding to the event sample reflected
in fig.~\ref{fig:t50} is shown in fig.~\ref{fig:s50}. 

It is instructive to regard the $\mathcal{S}$ distribution as being
composed mainly of a combinatorial background with in addition small "matching energy"
($\mathcal{E}$) and "matching position"($\mathcal{P}$) components. 
The meaning of these $\mathcal{E}$ and $\mathcal{P}$ components will be explained hereafter.\\
Consider the case that photon $k$ and photon $l$ are created by the decay of a certain $\pi^{0}$.
As a consequence, the combination $E_{k}\vec{x}_{k} \otimes E_{l}\vec{x}_{l}$ will then contribute
to the $\pi^{0}$ peak in the total invariant-mass distribution $\mathcal{T}$.
In the position-swapping procedure, the combination $E_{k}\vec{x}_{k} \otimes E_{l}\vec{x}_{m}$
will be encountered in the case that the randomly chosen photon is photon $m$ ($m \neq k \neq l$).
However, in this particular combination the energy $E_{l}$ happens to be the energy of the
photon which actually originated together with photon $k$ from the decay of one and
the same $\pi^{0}$.
The resulting entry in the invariant-mass distribution thus contains two photons with what
we call "matching energies" and consequently contribute to the $\mathcal{E}$ component.\\
In a similar way, the combination $E_{k}\vec{x}_{k} \otimes E_{m}\vec{x}_{l}$ contributes to the
$\mathcal{P}$ component in which both photon positions correspond to those of the two photons
originating from the $\pi^{0}$ decay.
Here the randomly chosen photon happens to be photon $l$, whereas the initial photon pair
consisted of photons $k$ and $m$.\\
The $\mathcal{E}$ and $\mathcal{P}$ components of the $\mathcal{S}$ distribution are intrinsically
different from the combinatorial background, as outlined below.
The remaining combinations entering the $\mathcal{S}$ distribution yield a distribution in
very good correspondance with the actual combinatorial background.

Consider the case that for some event $N_{phot}$ photons are detected and that the $\pi^{0}$
invariant-mass peak for this event contains $N_{peak}$ entries resulting from actual $\pi^{0}$ decays.
The combinatorial background in the total distribution $\mathcal{T}$ then consist of
$[\frac{1}{2}N_{phot}(N_{phot}-1)-N_{peak}]$ entries.
The position-swapped distribution $\mathcal{S}$ for this particular event has an $\mathcal{E}$ component 
with exactly $N_{peak}$ entries. The $\mathcal{P}$ component will, statistically, have $N_{peak}$
entries as well. The remaining $[\frac{1}{2}N_{phot}(N_{phot}-1)-2N_{peak}]$ entries in the
$\mathcal{S}$ distribution contribute to the combinatorial background.

It is crucial to accurately determine the $\mathcal{E}$ and $\mathcal{P}$ contributions
in order to determine the $\pi^{0}$ peak contents. First, the procedure to construct the $\mathcal{E}$ 
distribution will be described. It will be explained hereafter that in this process 
an additional distribution $\mathcal{U}$ is needed, consisting of entries of the form 
$\sqrt{\frac{1-\cos \theta _{A}}{1-\cos\theta _{B}}}$. Here $\theta _{A}$ and $\theta _{B}$
represent random angles between two photon momenta from the same event.
This distribution is normalised to have an integral equal to unity in order to ease
a proper scaling to the actual peak contents afterwards.

The invariant mass $M_{\mathcal{E}}$ of the combination $E_{k}\vec{x}_{k} \otimes E_{l}\vec{x}_{m}$
contributing to the $\mathcal{E}$ distribution, in which $m$ is the randomly selected photon,
is given by

\begin{equation}
M_{\mathcal{E}}=\sqrt{2E_{k}E_{l}(1-\cos \theta _{km})}=\sqrt{2E_{k}E_{l}(1-\cos
\theta _{kl})}\sqrt{\frac{(1-\cos \theta _{km})}{(1-\cos \theta _{kl})}}%
=M_{kl}\sqrt{\frac{(1-\cos \theta _{km})}{(1-\cos \theta _{kl})}}.
\end{equation}
As indicated by this formula, the $\mathcal{E}$ component of the $\mathcal{S}$ 
distribution is obtained by combination of the $\pi^{0}$ peak ($\mathcal{W}$) with the
$\mathcal{U}$ distribution in the following way~:

\begin{equation}
  \mathcal{E}(E)=\int_{0}^{\infty} {\rm d}u \int_{0}^{\infty} \mathcal{U}(u) 
  \mathcal{W}(E_{w}) \delta (E-uE_{w}) \,{\rm d}E_{w} ~~. 
\end{equation}

In a similar way we can construct a distribution that represents the $\mathcal{P}$ component
of the $\mathcal{S}$ distribution. For this a distribution $\mathcal{V}$ is needed that
is constructed from ratios of random energies, i.e. $\sqrt{\frac{E_{A}}{E_{B}}}$, as indicated
hereafter.\\
The invariant mass $M_{\mathcal{P}}$ of the combination $E_{k}\vec{x}_{k} \otimes E_{m}\vec{x}_{l}$
is~:

\begin{equation}
M_{\mathcal{P}}=\sqrt{2E_{k}E_{m}(1-\cos \theta _{kl})}=\sqrt{2E_{k}E_{l}(1-\cos
\theta _{kl})}\sqrt{\frac{E_{m}}{E_{l}}}=M_{kl}\sqrt{\frac{E_{m}}{E_{l}}}.
\end{equation}

Consequently, the $\mathcal{P}$ component of the $\mathcal{S}$ distribution is obtained by combination 
of the $\pi^{0}$ peak ($\mathcal{W}$) with the $\mathcal{V}$ distribution in the 
following way~:

\begin{equation}
  \mathcal{P}(E)=\int_{0}^{\infty} {\rm d}v \int_{0}^{\infty} \mathcal{V}(v) 
  \mathcal{W}(E_{w}) \delta (E-vE_{w}) \,{\rm d}E_{w} ~~. 
\end{equation}

We have now determined all the necessary components to construct the background of the
two-photon invariant-mass spectrum. This will allow us to extract the number of entries
in the $\pi^{0}$ peak due to genuine $\pi^{0}$ decays and thus the actual number of
detected $\pi^{0}$s within a certain event sample.

The distribution that results when the $\mathcal{S}$ distribution is
subtracted from the $\mathcal{T}$ distribution will be called the
"difference" distribution $\mathcal{D}$. It consists of the $N_{peak}$ 
entries in the $\pi ^{0}$ peak, $N_{peak}$ entries for the combinatorial 
background, minus the $\mathcal{E}$ and $\mathcal{P}$ components of the $\mathcal{S}$
distribution. An overview of all these contributions is provided in table~\ref{tabel1}.

\begin{table}[htb]
\begin{center}
\begin{tabular}{|l||l|l|l|l|}
\hline
&  $\pi^{0}$ peak & comb. background & $\mathcal{E}$ & $\mathcal{P}$\\
\hline 
$\mathcal{T}$ &  $N_{peak}$ &  $\frac{1}{2}N_{phot}(N_{phot}-1)-N_{peak}$ & - & - \\
\hline
$\mathcal{S}$& - &  $\frac{1}{2}N_{phot}(N_{phot}-1)-2N_{peak}$& $N_{peak}$ & $N_{peak}$ \\
\hline
$\mathcal{D}=\mathcal{T}-\mathcal{S}$ & $N_{peak}$ & $N_{peak}$ & $-N_{peak}$ & $-N_{peak}$\\
\hline
\end{tabular}
\end{center}
\caption{Composition of the various distributions in terms of the number of entries.}
\label{tabel1}
\end{table}

\begin{figure}[htb]
\begin{center}
\resizebox{11cm}{!}{\includegraphics{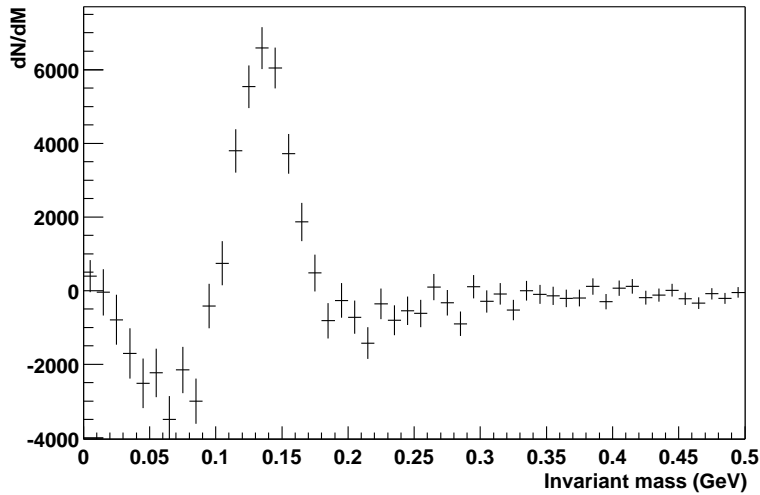}}
\end{center}
\caption{The $\mathcal{D}$ distribution for the case of 50 $\pi^{0}$s/event.}
\label{fig:d50}
\end{figure}

An example of this $\mathcal{D}$ distribution for the case of 50 $\pi^{0}$s/event
is shown in fig.~\ref{fig:d50}. It was obtained by subtracting 
the $\mathcal{S}$ distribution of fig.~\ref{fig:s50} from the $\mathcal{T}$ 
distribution of fig.~\ref{fig:t50}.

We now represent the actual $\pi^{0}$ peak by a gaussian of which the mean, standard deviation 
and the number of entries are regarded as free parameters. 
The $\mathcal{E}$ and $\mathcal{P}$ components corresponding to each of 
the trial gaussian distributions are obtained in the way described above. In this 
procedure the position-swapped distribution $\mathcal{S}$ provides an accurate
description of the combinatorial background. The four distributions $\mathcal{T}$, $\mathcal{S}$,
$\mathcal{E}$ and $\mathcal{P}$ are then combined to give a prediction for the $\mathcal{D}$
distribution, denoted by $\mathcal{D}_{pred}$. 

In order to obtain the best matching of the calculated ($\mathcal{D}_{pred}$) with the actual
$\mathcal{D}$ distribution, a $\chi^{2}$ value per bin is computed for each of the trial gaussian
distributions. This is performed in the $2\sigma$ interval around the $\pi^{0}$ mass. The reason 
that we restrict ourselves to the $2\sigma$ region for the fitting procedure is that in 
the tails the actual $\pi^{0}$ peak is not accurately described by a gaussian distribution,
which would have a degrading effect in the fitting procedure. The
gaussian distribution that produces the $\mathcal{D}_{pred}$ distribution that fits best to the
actual $\mathcal{D}$ distribution represents the actual $\pi^{0}$ peak in the interval
mentioned above. Consequently, a measure for the number of $\pi^{0}$s of which
the decay photons were both detected is obtained by performing the integral over the thus
obtained gaussian distribution.

\section{Results}
The method described above has been applied to the analysis of two sets of simulated data. 
Details concerning the simulation process will be given hereafter.\\
The first set consisted of the data of 50,000 events, where the number of $\pi^{0}$s produced per event
was equal to 50. This dataset corresponds to peripheral events as detected in the WA98 experiment.
Of the 100 decay photons per event, about 16 were detected within the acceptance of the WA98 photon
spectrometer \cite{wa98pt}. 

\begin{figure}[htb]
\begin{center}
\resizebox{12cm}{!}{\includegraphics{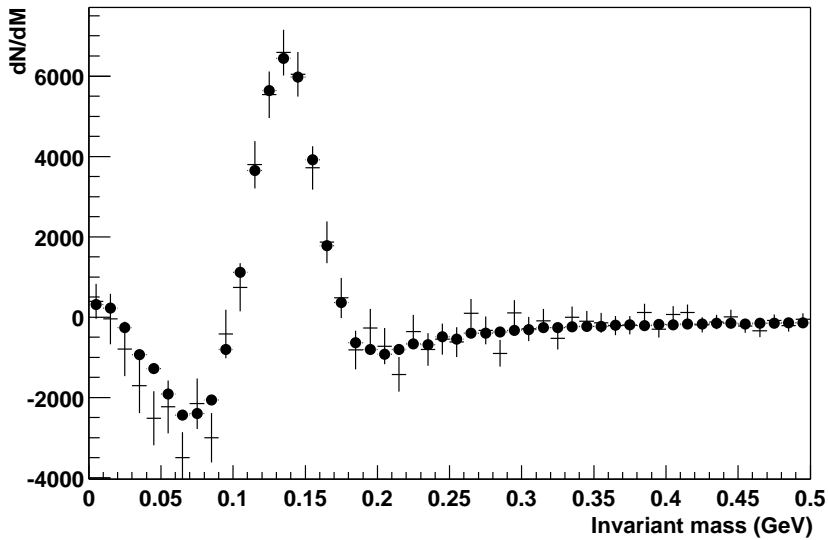}}
\end{center}
\caption{The $\mathcal{D}$ distribution for the case of 50 $\pi^{0}$s/event.
         The full circles indicate the best fit $\mathcal{D}_{pred}$.}
\label{fig:d50fit}
\end{figure}

In the fitting procedure of the $\mathcal{D}$ distribution, only the bins corresponding 
to the range of 90 to 180 MeV for the invariant mass were used. In fig.~\ref{fig:d50fit} 
the $\mathcal{D}$ distribution for this set of data is shown, together with the best 
fit $\mathcal{D}_{pred}$. 

\begin{figure}[htb]
\begin{center}
\resizebox{12cm}{!}{\includegraphics{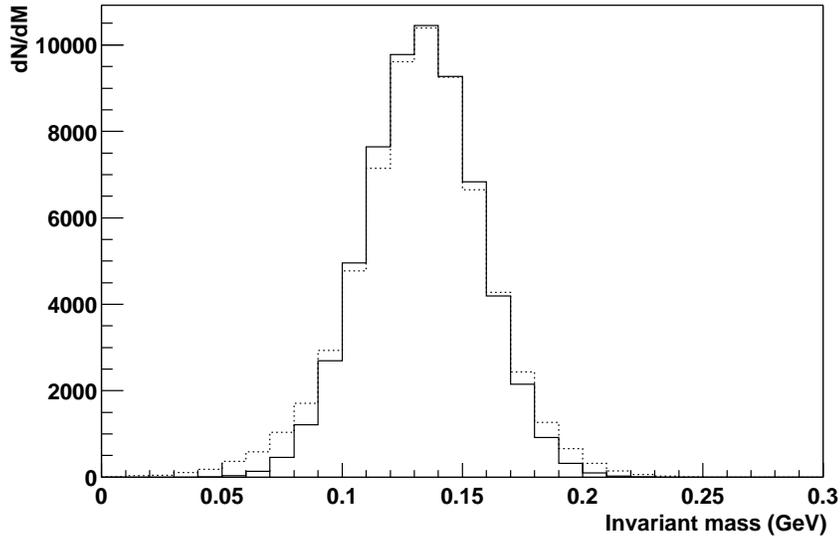}}
\end{center}
\caption{The input $\pi^{0}$ peak (dotted line) and the reconstructed $\pi^{0}$ peak
         for the case of 50 $\pi^{0}$s/event.}
\label{fig:pi50}
\end{figure}

In fig.~\ref{fig:pi50} the actual $\pi^{0}$ peak, as entered into the simulation, is 
presented as a dotted line, together with the reconstructed gaussian for the $\pi^{0}$ peak
determined from the best fit for the reconstructed $\mathcal{D}$ distribution.
The integral of the input $\pi^{0}$ peak over the fitted interval is 57,479, whereas the result
obtained with our reconstruction method amounts to 57,977.

\newpage

\begin{figure}[htb]
\begin{center}
\resizebox{12cm}{!}{\includegraphics{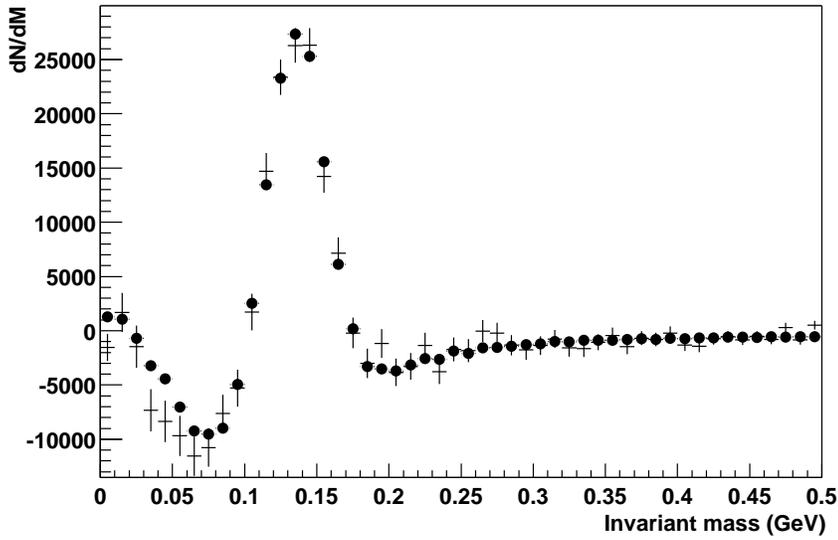}}
\end{center}
\caption{The $\mathcal{D}$ distribution for the case of 100 $\pi^{0}$s/event.
         The full circles indicate the best fit $\mathcal{D}_{pred}$.}
\label{fig:d100fit}
\end{figure}

The second dataset consisted of 100,000 events. The number of $\pi^{0}$s per event
was 100, corresponding to medium central events as detected in WA98. 
Of the 200 decay photons per event, on average about 32 were detected. 
Fig.~\ref{fig:d100fit} shows the $\mathcal{D}$ distribution for this set of data, together 
with the best fit $\mathcal{D}_{pred}$. 

\begin{figure}[htb]
\begin{center}
\resizebox{12cm}{!}{\includegraphics{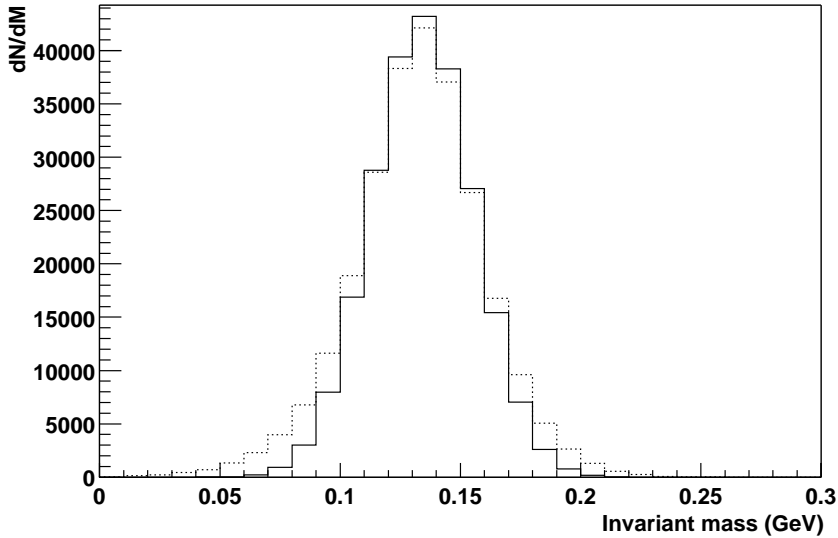}}
\end{center}
\caption{The input $\pi^{0}$ peak (dotted line) and the reconstructed $\pi^{0}$ peak
         for the case of 100 $\pi^{0}$s/event.}
\label{fig:pi100}
\end{figure}

Fig.~\ref{fig:pi100} shows the input $\pi^{0}$ peak (dotted line), together with 
the reconstructed gaussian for the $\pi^{0}$ peak. The integral of the input $\pi^{0}$ peak 
over the fitting interval is 229,712, whereas our reconstruction yields 224,023.\\
It is seen that in both cases our reconstruction results are in very good agreement with the
number of $\pi^{0}$s as entered through the computer simulation. 

\section{The computer simulation procedure}
\label{sec-compsim}
In the computer simulation that is used to test the method, a fixed number of 
$\pi ^{0}$s is created for each event. All of the $\pi ^{0}$s were decayed into 
a pair of photons.\\
In order to obtain an exponential-like spectrum as observed 
experimentally for various particle species \cite{prl85,na44,na49}, the $\pi^{0}$ energies
were distributed according to a Bose-Einstein distribution corresponding to a temperature
of 200~MeV.\\
The amount of $\pi^{0}$s was chosen to match realistic conditions for high-energy heavy ion experiments.
The number of accepted photons is comparable to the number of detected photons in 
peripheral or medium central events in the WA98 experiment.

To mimic realistic experimental conditions, the directions of the photon momenta were compared to the 
acceptance angles of the WA98 lead-glass detector and only the photons within the 
acceptance were used in our analysis. In order to obtain a dataset that models the actual experimental 
data in a realistic way, the imperfections of the detector were simulated as well. 
The energy resolution of the detector, given by $\frac{\sigma}{E}=0.02+\frac{0.08}{\sqrt{E}}$ where
all energies are in GeV, was used to obtain a smearing of the photon energies.
Furthermore, we randomly discard 10\% of all photons to simulate the effect of efficiency loss
which in WA98 is due to charged particle vetoing. 

\section{Summary}
The method introduced in this report allows accurate determination of the number of 
$\pi^{0}$s via their two-photon decay channel in a range of 2$\sigma$ around the 
$\pi^{0}$ mass.
The procedure does not invoke event-mixing techniques but is applied on an event-by-event
basis using the difference between the total invariant-mass distribution ($\mathcal{T}$) and a 
newly introduced "position swapped" distribution $\mathcal{S}$.

The method has been tested by means of computer simulations modelling peripheral
and medium central events in Pb+Pb collisions at CERN-SPS energies.
For these peripheral and medium central event samples it was seen that the $\pi^{0}$
yield could be extracted with an accuracy of 0.9\% and 2.5\%, respectively.
As a final remark we would like to mention that the correlations of the $\mathcal{T}$ and
$\mathcal{S}$ distributions have not been taken into account in the computation of the
uncertainties of the "difference" distributions $\mathcal{D}$.
However, it was seen that in the final results the effect of these correlations was negligible.\\
In addition, we believe that this method, with some modifications, could be applied to the analysis of 
other two-particle decays, e.g. charm measurement via the $D^{0} \rightarrow K\pi$ decay channel.

\begin{ack}
The authors would like to thank Eug\`{e}ne van der Pijll and Garmt de Vries
for the very fruitful discussions on the subject.
\end{ack}

\end{document}